\documentclass[aps,prl,twocolumn,pacs]{revtex4-1}
\usepackage{amsmath,bm,epsfig}
\usepackage{epsfig}
\usepackage{bm}
\newcommand{\B}[1]{{\bm{#1}}}
\newcommand{\C}[1]{{\mathcal{#1}}}
\newcommand{\pa}{\partial}
\usepackage[latin1]{inputenc}
\usepackage{graphicx}

\begin{document}
\title{A nonlinear symmetry breaking effect in shear cracks}
\author{Roi Harpaz and Eran Bouchbinder}
\affiliation{Chemical Physics Department, Weizmann Institute of Science, Rehovot 76100, Israel}
\begin{abstract}
Shear cracks propagation is a basic dynamical process that mediates interfacial failure. We develop a general weakly nonlinear elastic theory of shear cracks and show that these experience tensile-mode crack tip deformation, including possibly opening displacements, in agreement with Stephenson's prediction. We quantify this nonlinear symmetry breaking effect, under two-dimensional deformation conditions, by an explicit inequality in terms of the first and second order elastic constants in the quasi-static regime and semi-analytic calculations in the fully dynamic regime. Our general results are applied to various materials. Finally, we discuss available works in the literature and note the potential relevance of elastic nonlinearities for frictional cracks.
\end{abstract}
\maketitle

A fundamental process that governs the strength and failure of interfaces in a wide range of scientific and technological problems is the propagation of shear cracks. For example, stress and energy release during earthquakes is mediated by the propagation of shear cracks along fault surfaces \cite{02Scholz}. In generic situations, shear driven cracks (also known as mode II symmetry cracks \cite{98Fre}) involve interfacial slip, but no crack tip opening displacements and/or tensile stresses, which are characteristic of mode I symmetry cracks \cite{98Fre}. However, the existence of such a symmetry breaking effect is of interest and potential importance. This becomes evident when the propagation of shear cracks is accompanied by frictional processes. Under such circumstances, crack opening, or even a reduction in the confining normal stress at the interface, can significantly reduce the friction force and hence result in interface weakening.

The most widely invoked mechanism for the coupling between interfacial slip and normal stress variations (and possibly crack opening) is material discontinuity, i.e. the shear crack is assumed to propagate along an interface separating two dissimilar materials \cite{01BZ}. An alternative physical mechanism, operative even in the absence of material discontinuity, is elastic nonlinearity \cite{71Schallamach, 82Steph}.

In this Letter, we explore the latter mechanism by deriving general weakly nonlinear asymptotic solutions for frictionless shear cracks in homogeneous isotropic materials. The conditions for crack tip opening, as a function of the first and second order elastic constants and crack propagation velocity, are quantified and the theory is applied to polymeric, glassy and metallic materials. Finally, we discuss the possible relevance of our results to frictional shear cracks and to the super-shear transition \cite{00AG}.

We note that while the absence of pure shear crack solutions in nonlinear elastic fracture mechanics was recognized by Stephenson in 1982 \cite{82Steph}, to the best of our knowledge no accurate and self-consistent benchmark weakly nonlinear solutions are presently available (see below).

We start by considering the motion $\B \varphi$, which is assumed to be a continuous, differentiable
and invertible mapping between a reference configuration $\B x$ and
a deformed configuration $\B x'$ such that $\B x'\!=\!\B \varphi(\B
x)\!=\!\B x\!+\!\B u(\B x)$, where $\B u(\B x)$ is the displacement field. The Green-Lagrange (metric) strain tensor $\B E$
is defined in terms of the deformation gradient tensor $\B F \!=\! \nabla\B\varphi\!=\!\B I \!+\! \B H$ as $\B E \!=\! \case{1}{2}(\B F^T \B F \!-\!\B I)$ \cite{51Murnaghan}. Here $\B I$ is the
identity tensor and $\B H\!=\!\nabla \B u$ is the displacement gradient tensor.

The elastic energy per unit reference volume $U(\B E)$ of isotropic and compressible materials can be expressed in terms of the scalar invariants of $\B E$, which are conveniently written as $tr\!\B E$, $tr\!\B E^2$ and $tr\!\B E^3$.
To third order in $\B E$, $U(\B E)$ takes the form
\begin{eqnarray}
\label{U_E}
&&U(\B E) = \frac{\lambda}{2} (tr\!\B E)^2 + \mu\,tr\!\B E^2 + \\
&&~~~\frac{\alpha_1}{3} (tr\!\B E)^3 + \alpha_2 \,tr\!\B E~ tr\!\B E^2 + \frac{2\alpha_3}{3}tr\!\B E^3 + \C O(\B E^4) \ ,\nonumber
\end{eqnarray}
where $\{\lambda, \mu\}$ are the first order (Lam\'e) constants and $\{\alpha_i\}_{i=1-3}$ are second order elastic constants. The latter are important physical constants that are closely related to the leading anharmonic contributions to interatomic interaction potentials. These anharmonic contributions are known to be the origin of many basic physical phenomena and properties such
as the Gr\"{u}neisen parameters, deviations from the Dulong-Petit law at high temperatures, the thermal expansion coefficient and the existence of thermal resistance. Their implications for materials failure
remain largely unexplored (but see \cite{81Knowles, 86COL, 04CHGW, 08LBF, 08BLF, 09BLF, 10BLIF, 10Bou}), a situation that the present Letter aims to at least partially improve.

For simplicity, in what follows we focus on plane strain conditions \cite{comment} for which Eq. (\ref{U_E}) reduces to
\begin{eqnarray}
\label{U_E_2d}
\hspace{-0.4cm}U^{2D}(\B E) = \frac{\lambda}{2} (tr\!\B E)^2 + \mu\,tr\!\B E^2 + \frac{l}{3} (tr\!\B E)^3 +\frac{2m}{3}tr\!\B E^3  \ ,
\end{eqnarray}
where $\B E$ is understood hereafter as a 2D tensor and $l\!=\!\alpha_1\!+\!\alpha_2$ and $m\!=\!\alpha_3\!+\!\alpha_2$ are two of the Murnaghan coefficients \cite{51Murnaghan}.

To develop a fracture theory based on this general expansion we write down the momentum balance equation
\begin{equation}
\label{EOM}
\nabla \cdot \B s = \rho \,\pa_{tt}{\B \varphi} \ ,
\end{equation}
where $\rho$ is the reference mass density and $\B s\!=\!\pa_{\B F} U^{2D}$ is the first Piola-Kirchhoff stress
tensor, that is work-conjugate to the deformation gradient $\B F$ \cite{51Murnaghan}. A frictionless crack is introduced by the usual
traction-free boundary conditions
\begin{equation}
\label{BC}
s_{xy}(r,\theta\!=\!\pm\pi)\!=\!s_{yy}(r,\theta\!=\!\pm\pi)=0 \ ,
\end{equation}
where $(r,\theta)$ is a polar coordinate system that moves with the
crack tip and is related to the reference frame by
$r\!=\!\sqrt{(x-vt)^2+y^2}$ and $\theta\!=\!\tan^{-1}[y/(x-vt)]$. $x$ is the propagation direction, $y$ is the perpendicular direction and $v$ is the tip propagation velocity. In what follows
we neglect crack acceleration effects, implying that all of the fields depend on $x$ and $t$
through the combination $x\!-\!vt$ and therefore $\pa_t\!=\!-v\pa_x$.

To proceed, we expand $\B s$ order by order in powers of the displacement gradient $\B H$ to obtain
\begin{eqnarray}
\label{s_2nd_order}
\!\!\!\!\!\!&&\B s(\B H) \simeq \lambda~\! tr\B \varepsilon \B I+ 2 \mu \B \varepsilon + \frac{1}{2} \lambda~ tr(\B H^T\B H) \B I + \mu\, \B H^T \B H+\nonumber\\
\!\!\!\!\!\!&&\lambda \,tr \B \varepsilon \B H + 2 \mu\, \B H \B \varepsilon + l \, (tr\B \varepsilon)^2 \B I + 2 m\, \B \varepsilon^2 + \C O(\B H^3)\ ,
\end{eqnarray}
where $\B \varepsilon \!=\! \case{1}{2}(\B H + \B H^T)$ is the infinitesimal strain tensor.
We then follow \cite{81Knowles, 86COL, 04CHGW, 08LBF, 08BLF, 09BLF, 10BLIF, 10Bou, 65TN} and write the displacement field as
\begin{equation}
\label{expansion}
\B u(r,\theta) \simeq \B u^{(1)}(r,\theta)+ \B u^{(2)}(r,\theta) \ ,
\end{equation}
where the superscripts denote different orders in $\B H$. Substituting Eqs. (\ref{s_2nd_order})-(\ref{expansion}) into Eqs. (\ref{EOM})-(\ref{BC}), we obtain the equations and boundary
conditions for $\B u^{(1)}$ and $\B u^{(2)}$.

To linear order, we obtain the well-studied equations of isotropic Linear Elastic Fracture Mechanics (LEFM) \cite{98Fre}.
To second order, we obtain
\begin{equation}
\mu\nabla^2{\B u^{(2)}}+(\lambda+\mu)\nabla(\nabla\cdot{\B u^{(2)}})+
\B {\C F}[\B u^{(1)}]=\rho\ddot{\B u}^{(2)}\ ,
\label{secondO}
\end{equation}
where the boundary conditions at $\theta\!=\!\pm\pi$ read
\begin{eqnarray}
\label{BC2nd}
&&-r^{-1}\pa_\theta u_x^{(2)}-\pa_r u_y^{(2)}-\C S_x[\B u^{(1)}] = 0\ , \nonumber\\
&&-(\lambda+ 2\mu)\, r^{-1} \pa_\theta u_y^{(2)}-\lambda \,\pa_r u_x^{(2)}-\C S_y[\B u^{(1)}]  = 0 \ .
\end{eqnarray}
The functionals $\B {\C F}[\B u^{(1)}]$ and $\B {\C S}[\B u^{(1)}]$ are easily obtained from Eq. (\ref{s_2nd_order}). Asymptotic near-tip solutions of $\B u^{(2)}$ for mode I (tensile) symmetry were derived in \cite{08BLF, 09BLF, 10BLIF, 10Bou}, with a focus on a particular constitutive law.

Here we focus on global mode II (shear) symmetry loading. Under these conditions, the linear order solution $\B u^{(1)}$ possesses the following symmetry properties
\begin{eqnarray}
\label{symmetry II}
\!\!\!\!\!\!u_x^{(1)}(r,-\theta)\!=\! -u_x^{(1)}(r,\theta)~~\hbox{and}~~ u_y^{(1)}(r,-\theta) \!=\! u_y^{(1)}(r,\theta).
\end{eqnarray}
The leading order near-tip asymptotic solution reads \cite{98Fre}
\begin{eqnarray}
\label{u_1st_II}
&&\!\!\!u_x^{(1)}(r, \theta) = \frac{2K_{II}}{\mu \sqrt{2\pi}D(v)} \times \\
&&~~~\left[2\alpha_s r_d^{1/2}\sin{\left(\theta_d/2\right)}-\alpha_s(1+\alpha_s^2) r_s^{1/2}\sin{\left(\theta_s/2\right)}\right],\nonumber\\
&&\!\!\!u_y^{(1)}(r,\theta) = \frac{2K_{II}}{\mu \sqrt{2\pi}D(v)} \times \nonumber\\
&&~~~\left[2\alpha_s\alpha_d r_d^{1/2}\cos{\left(\theta_d/2\right)}-(1+\alpha_s^2)r_s^{1/2}\cos{\left(\theta_s/2\right)} \right] \nonumber\ .
\end{eqnarray}
Here $\alpha^2_{d,s} \!=\! 1\!-\!v^2/c_{d,s}^2$, $\tan{\theta_{d,s}}\!=\!\alpha_{d,s}\tan{\theta}$, $\quad r_{d,s}\!=\!r\sqrt{1\!-\!(v\sin\theta/c_{d,s})^2}$, $D(v)\!=\! 4\alpha_s\alpha_d\!-\!(1\!+\!\alpha_s^2)^2$; $K_{II}$ is the mode II ``stress intensity factor'' and $c_d\!=\!\sqrt{(\lambda+2\mu)/\rho},~~ c_s\!=\!\sqrt{\mu/\rho}$ are the dilatational and shear wave speeds, respectively \cite{98Fre}.

Using Eqs. (\ref{u_1st_II}) to calculate $\B {\C F}[\B u^{(1)}]$ and $\B {\C S}[\B u^{(1)}]$ in Eqs. (\ref{secondO})-(\ref{BC2nd}), we obtain                                                                                                                  \begin{eqnarray}
\label{F_S}
\B{\C F}^{II}(r,\theta) &=& \frac{K^2_{II} \B g^{II}(\theta;v)}{\mu^2 r^2}, ~~\C S^{II}_x(r,\pm\pi) = 0 \ ,\nonumber\\
\C S^{II}_y(r,\pm\pi) &=&  \frac{K^2_{II} \kappa^{II}(v)}{\mu^2 r} \ ,
\end{eqnarray}
where $\B g^{II}(\theta;v)$ and $\kappa^{II}(v)$ are functions of stress dimension that are too lengthy to report here (the superscript denotes that these quantities correspond to mode II loading conditions). $\B{\C F}^{II}(r,\theta)$ and $\B {\C S}^{II}_y(r,\pm\pi)$, when substituted into Eqs. (\ref{secondO})-(\ref{BC2nd}), can be respectively interpreted as an effective body force and a surface force in an effective linear elastic crack problem.
Their symmetry properties dictate the symmetry properties of the weakly nonlinear solution $\B u^{(2)}$. Equation (\ref{F_S}) indicates that $\B {\C S}^{II}(r,\pi)\!=\!\B {\C S}^{II}(r,-\pi)$ and an explicit calculation shows that
\begin{equation}
g^{II}_x(\theta;v)=g^{II}_x(-\theta;v)~~\hbox{and}~~ g^{II}_y(\theta;v)=-g^{II}_y(-\theta;v) \ .
\label{g_symmetry}
\end{equation}
These properties imply that the weakly nonlinear problem is of {\em pure} mode I nature, even though the loading is of pure mode II nature \cite{81Knowles, 86COL, 04CHGW}. Therefore, while $\B u^{(1)}$ possesses mode II symmetry, cf. Eq. (\ref{symmetry II}), $\B u^{(2)}$ satisfies
\begin{eqnarray}
\label{symmetry I}
\!\!\!\!\!\!\!\!\!u_x^{(2)}(r,-\theta) \!=\! u_x^{(2)}(r, \theta) ~\hbox{and}~ u_y^{(2)}(r,-\theta) \!=\! -u_y^{(2)}(r, \theta),
\end{eqnarray}
which is the nonlinear symmetry breaking effect we aim at studying here.

To quantify the effect we need to solve the problem posed in Eqs. (\ref{secondO}), (\ref{BC2nd}) and (\ref{F_S}). Since this is a pure mode I problem, we can follow \cite{08BLF, 09BLF, 10BLIF, 10Bou} to obtain
\begin{widetext}
\begin{eqnarray}
\label{solution}
u_x^{(2)}(r,\theta) &=& \frac{K^2_{II}}{\mu^2}\left[A \log{r} + \frac{A}{2}\log{\left(1-\frac{v^2\sin^2\theta}{c_d^2} \right)} + B\alpha_s\log{r} + \frac{B
\alpha_s}{2}\log{\left(1-\frac{v^2\sin^2\theta}{c_s^2} \right)}+\Upsilon^{II}_x(\theta;v)\right],\nonumber\\
u_y^{(2)}(r,\theta)&=&\frac{K^2_{II}}{\mu^2}\Big[-A\alpha_d\theta_d-B\theta_s+\Upsilon^{II}_y(\theta;v)\Big]\ .
\end{eqnarray}
\end{widetext}
$\B \Upsilon^{II}(\theta;v)$ is an $r$-independent solution of Eq. (\ref{secondO}) that does not satisfy the boundary conditions in Eq. (\ref{BC2nd}). It can readily obtained semi-analytically using a Fourier series method \cite{08BLF, 09BLF}.
The rest of the solution satisfies Eq. (\ref{secondO}) when the effective body force is omitted. It contains two parameters, $A$ and $B$, that should be determined by the boundary conditions. However, the first equation in
(\ref{BC2nd}) is satisfied automatically and only the second one imposes a constraint of the form
\begin{equation}
A = \frac{2\mu B \alpha_s -(\lambda+2\mu)\pa_\theta \Upsilon^{II}_y(\pi;v)-\kappa^{II}(v)}{\lambda - (\lambda+2\mu)\alpha_d^2} \ .
\label{A_B_bc}
\end{equation}

To complete the solution, we need an additional condition that allows the calculation of both $A$ and $B$. To see where this additional condition emerges from, substitute the solution in Eq. (\ref{solution}) in $\B s$ of Eq. (\ref{s_2nd_order}) and write $\B s =  \B s^{(1)} \!+\! \B s^{(2)}$, where the superscripts denote different orders of $K_{II}$. $\B s^{(2)}$ is characterized by a $1/r$ spatial singularity that was shown to give rise to a (spurious) force in the crack parallel direction, $f^{(2)}_x$, a force that cannot be balanced by material inertia which vanishes to this order \cite{09BLF, 10Bou, 74Rice}. Therefore, we need to impose the condition
\begin{eqnarray}
\label{dynamic_condition}
f^{(2)}_x = \int_{-\pi}^{\pi} s^{(2)}_{xj} n_j r d\theta = 0 \ ,
\end{eqnarray}
ensuring that our solution is consistent with the conservation of momentum, which is otherwise violated. Equation (\ref{dynamic_condition}), which is a linear relation between $A$ and $B$, together with Eq. (\ref{A_B_bc}), allows us to fully determine the solution in Eq. (\ref{solution}).

A natural and physically transparent way to quantify the symmetry breaking effect is through the dimensionless crack tip displacement defined as
\begin{eqnarray}
\label{Delta}
\Delta \equiv  \frac{\mu^2 \left[u^{(2)}_y (r,\pi)-u^{(2)}_y (r,-\pi)\right]}{K^2_{II}} \ .
\end{eqnarray}
Crack tip opening emerges when $\Delta\!>\!0$. The complementary situation, $\Delta\!<\!0$, obviously does not imply that the crack's faces penetrate each other, but rather that the traction-free boundary conditions are no longer valid and other physical effects should be taken into account.
Equation (\ref{solution}) immediately tells us that $\Delta$ is $r$-independent. Hence, we are mainly interested in the dependence of $\Delta$ on the first and second order elastic constants and on the crack velocity, i.e. $\Delta(\lambda, \mu, l, m, v)$. To study this function we first focus on the quasi-static limit $v\!\to\!0$, and later examine the behavior in the fully dynamic regime.

The problem for $v\!=\!0$ can be solved analytically and leads to
\begin{eqnarray}
\label{Delta static solution}
&&\hspace{-1cm}\Delta(v\!=\!0)=\\
&&\hspace{-1cm}\frac{(1-2\nu)^2}{4}\!+\!\frac{(1-2\nu)^3}{4}\frac{l}{\mu}\!-\!\frac{\nu(1-\nu)(1-2\nu)}{2}\frac{m}{\mu} \ ,\nonumber
\end{eqnarray}
where $\nu\!=\!\lambda/[2(\lambda\!+\!\mu)]$ is Poisson's ratio. We first note that $\Delta$ depends only on three dimensionless material parameters, $\nu$, $l/\mu$ and $m/\mu$. In addition, $\Delta$ contains
a common factor inversely proportional to the bulk modulus $K\!\sim\!(1-2\nu)^{-1}$ and may possibly feature an interesting behavior as the incompressibility limit is approached, $\nu\!\to\!1/2$. We are mainly interested in the sign of $\Delta$, and if positive, in its magnitude.
Therefore, Eq. (\ref{Delta static solution}) can in fact serve as an explicit inequality that $\nu$, $l/\mu$ and $m/\mu$ should satisfy in order to observe crack tip opening, $\Delta\!>\!0$.

In order to better understand this inequality, let us consider the different terms in Eq. (\ref{Delta static solution}). The first term is independent of the second order elastic constants $l$ and $m$, and hence is associated with a geometric nonlinearity. This term exists even in the absence of constitutive nonlinearities, $l\!=\!m\!=\!0$, and emerges because $\B E$ is nonlinear in $\B H$, see above. Most importantly, it is positive independently of $\nu$. Therefore, we conclude that geometric nonlinearities universally tend to induce opening of shear cracks in the quasi-static regime, $v\!=\!0$. The last two terms on the right-hand-side of Eq. (\ref{Delta static solution}) depend on constitutive nonlinearities quantified by $l$ and $m$. Since $l$ and $m$ can be both positive and negative (typically $l,m\!\gg\!\mu$), we cannot say something definite about these contributions and whether the inequality is satisfied or not depends on the specific values for each material.
\begin{figure}[here]
\centering \epsfig{width=0.55\textwidth,file=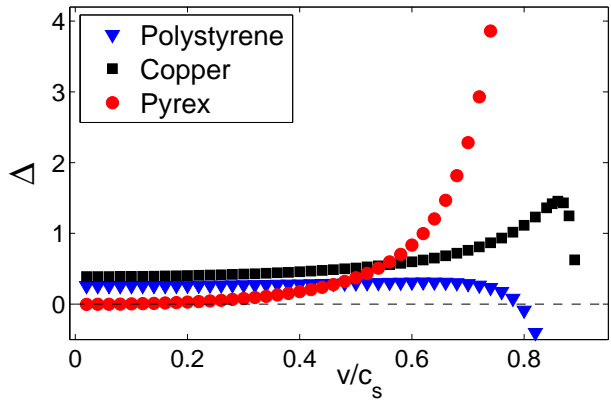}
\caption{$\Delta$ as a function of $v/c_s$ for three materials.} \label{fig1}
\end{figure}

It would be interesting to analyze the dependence of $\Delta$ on the crack propagation velocity $v$. To extend Eq. (\ref{Delta static solution}) to the dynamic regime we should introduce another dimensionless parameter, $v/c_s$, and write
\begin{eqnarray}
\label{Delta dynamic solution}
\hspace{-0.6cm}\Delta(v)\!=\!C_0(\nu,v/c_s)\!+\!C_l(\nu,v/c_s)\frac{l}{\mu}\!+\!C_m(\nu,v/c_s)\frac{m}{\mu} \ ,
\end{eqnarray}
where $C_0$, $C_l$ and $C_m$ are dimensionless functions that reduce to their counterparts in Eq. (\ref{Delta static solution}) when $v\!\to\!0$.
$\Delta(v)$ was calculated semi-analytically following the procedure described in \cite{08BLF, 09BLF, 10BLIF, 10Bou}. In Fig. \ref{fig1} we show $\Delta(v)$ for a polymer (Polystyrene with $\nu\!=\!0.34$, $l/\mu\!=\!-13.7$ and $m/\mu\!=\!-9.6$ \cite{53HK}), a glass (Pyrex with $\nu\!=\!0.17$, $l/\mu\!=\!0.5$ and $m/\mu\!=\!3.4$ \cite{53HK}) and a metal (Copper with $\nu\!=\!0.35$, $l/\mu\!=\!11.8$ and $m/\mu\!=\!-8.1$ \cite{66Crecraft}).  For Polystyrene, $\Delta(v)$ remains nearly $v$-independent until it changes sign at high $v$. For Pyrex, $\Delta(v)$ increases monotonically and quite significantly at high $v$. For Copper, $\Delta(v)$ increases up to a high $v$ where it abruptly decreases and eventually becomes negative. The interplay between materials parameters and elastodynamic effects gives rise to this rich behavior.
\begin{figure}[here]
\centering \epsfig{width=0.51\textwidth,file=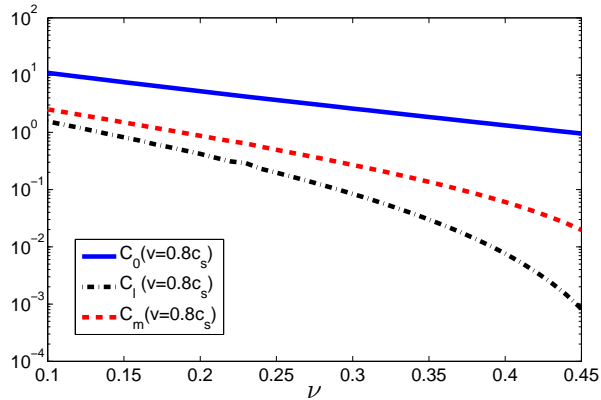}
\caption{A linear-log plot of the coefficient functions $C_0, C_l, C_m$, defined in Eq. (\ref{Delta dynamic solution}), vs. $\nu$ for $v\!=\!0.8c_s$.} \label{fig2}
\end{figure}

The extreme elastodynamic limit, $v\!\sim\!c_s$, is of particular interest. This regime might be relevant for phenomena such as ordinary earthquakes and the super-shear transition \cite{00AG}. To better understand what determines the sign of $\Delta$ at high $v$, we present in Fig. \ref{fig2} $C_0,\;C_l,\;C_m$ as a function of $\nu$ for $v\!=\!0.8c_s$. Several qualitative conclusions can be drawn. All $C$'s are positive and satisfy $C_0\!\gg\!C_m\!>\!C_l$, where $C_m\!\gg\!C_l$ for large values of $\nu$ ($\nu\!>\!0.3$). Therefore, $l$ typically plays a minor role in determining the sign of $\Delta(v\!=\!0.8c_s)$. In addition, positive values of $m$ and $l$ immediately ensure $\Delta(v\!=\!0.8c_s)\!>\!0$, which is the case of Pyrex, cf. Fig. \ref{fig1}. Finally, since $C_0$ is an order of magnitude (or more) larger than $C_m$, a large and negative $m$ is needed in order to have $\Delta(v\!=\!0.8c_s)\!<\!0$. This is the case of Polystyrene, cf. Fig. \ref{fig1}.

To put our results in the context of existing literature, we note that there were very few attempts to derive specific weakly nonlinear crack tip solutions \cite{81Knowles, 86COL, 04CHGW}. Invariably, all of these attempts {\em assumed} that
$\B u^{(2)}(r,\theta)$ is $r$-independent. Equations (\ref{solution}) clearly demonstrate that this is not the case as $u^{(2)}_x(r,\theta)$ contains a contribution proportional to $\log{r}$, which was experimentally verified for mode I fracture \cite{08BLF, 09BLF, 10BLIF}. More generally, Eqs. (\ref{solution}) show that $\B u^{(2)}(r,\theta)$ cannot be separated into a product of an $r$-dependent function and a $\theta$-dependent function. In essence, the assumption of an $r$-independent $\B u^{(2)}(r,\theta)$ amounts to implicitly adopting the condition $A + B \alpha_s = 0$, which annihilates the $\log{r}$ term in Eqs. (\ref{solution}). With this conditions at hand, indeed a solution of Eqs. (\ref{secondO}), (\ref{BC2nd}) and (\ref{F_S}) can be found. Unfortunately, such a solution {\em does not} satisfy Eq. (\ref{dynamic_condition}). Therefore, the solutions discussed in \cite{81Knowles, 86COL, 04CHGW} produce an unphysical force in the crack propagation direction and hence violate the conservation of momentum.

We have verified the latter statement in relation to the solution presented in \cite{86COL}, which was derived for a nonlinear Hookean material ($l\!=\!m\!=\!0$ in Eq. (\ref{U_E_2d})) in the quasi-static limit ($v \!\to\! 0$). We recovered this solution using $A\!+\!B\!=\!0$ (recall that $\alpha_s\!=\!1$ for $v\!=\!0$) instead of Eq. (\ref{dynamic_condition}). However, as explained above, this solution violates Eq. (\ref{dynamic_condition}), i.e. it produces a finite $f_x^{(2)}$. In addition, the solution in \cite{86COL} predicts crack closure, $\Delta<0$, which is inconsistent with Eqs. (\ref{Delta static solution}) that predicts $\Delta\!>\!0$ for $l\!=\!m\!=\!0$.

The scale of the opening displacement $\Delta$ and the distance from the crack tip in which it takes place are inherited from the lengthscale $K_{II}^2/\mu^2$. For example, for a material with $\mu\!=\!1$\,GPa and $K_{II}\!=\!5$\,MPa$\sqrt{m}$ (which may be relevant for polymers) we obtain a lengthscale of $\sim 25\mu$m. The value of $K_{II}$ (the dynamic mode II fracture toughness) is a material parameter that is, unfortunately, not well documented for many materials under dynamic conditions. All of our results can be recast in a dimensional form once $K_{II}$ is known.

The symmetry breaking effect we have quantitatively addressed in this work may have physical consequences for several important problems. As was already mentioned in the introduction, this effect may influence the strength of frictional interfaces. Since the local frictional resistance along interfacial contact depends linearly on the compressive normal stress, crack tip opening or even a reduction in this stress, may locally weaken the interface. Elastic nonlinearities, which are rather generally overlooked in frictional analysis, may provide a mechanism for such a local weakening near the crack tip. A quantitative analysis of this possible effect goes beyond the present work, entailing the introduction of a friction law acting on the crack faces and the inclusion of a compressive normal stress.

Another relevant problem is the super-shear transition beyond which a shear crack propagates at a velocity larger than the shear wave speed \cite{04Xia}. The transition from a sub-shear state to a super-shear state is believed to be mediated by the nucleation of a ``daughter'' crack in front of the ``mother'' crack \cite{00AG}. Above, we quantified the symmetry breaking effect as manifested at the crack faces, i.e. $\theta\!=\!\pm\pi$. However, tensile components of the displacement and stress fields emerge also at a continuous range of other angles, including ahead of the crack tip, potentially facilitating the super-shear transition. Indeed, the numerical simulations of \cite{00AG}, which demonstrated the super-shear transition, also feature opening displacement that might be interpreted as the symmetry breaking effect we addressed \cite{04CHGW}.

We hope that the present work, beyond providing general and self-consistent weakly nonlinear shear crack solutions, will serve as an impetus for more systematic studies of the roles played by elastic nonlinearities in the dynamics of shear cracks in various of problems.

{\bf Acknowledgements} This work was supported by the James S. McDonnell Foundation, the Harold Perlman Family Foundation and the William Z. and Eda Bess Novick Young Scientist Fund.


\begin{thebibliography}{99}
\bibitem{comment} We repeated our calculations under plane stress conditions and found no qualitative change in the results.
\end{thebibliography}
\end{document}